\documentclass[prb,showpacs,preprint,nobibnotes]{revtex4}
\usepackage{graphicx}
\begin{document}
\bibliographystyle{apsrev}

\title{Comparison of variational real-space representations of the kinetic energy operator }

\author{Chris-Kriton Skylaris}
\author{Oswaldo Di\'{e}guez}
\thanks{Corresponding author. Fax: +44 (0)1223 337356} 
\email{odl22@phy.cam. ac.uk}
\homepage{http://www.tcm.phy.cam.ac.uk/~odl22/} 
\author{Peter D. Haynes}
\author{Mike C. Payne}
\affiliation{Theory of Condensed Matter, Cavendish Laboratory, \\ 
Madingley Road, Cambridge CB3 0HE, UK }

\date{\today}

\begin{abstract}
We present a comparison of real-space  methods 
based on regular grids for 
electronic structure calculations that are 
designed to have basis set variational properties, 
using as a reference the conventional method of 
finite differences (a real-space method that is not variational)
and the reciprocal-space plane-wave method 
which is fully variational.
We find that a new definition of the finite differences 
method [P. Maragakis, J. Soler, and E. Kaxiras, Phys. Rev. B 
\textbf{64}, 193101 (2001)]
satisfies one of the two properties of variational
behaviour at the cost of larger errors than the conventional
finite differences method. On the other hand, a 
 technique which represents functions in a number 
of plane-waves which is independent of system size,
 follows closely  the 
plane-wave method and therefore also the
criteria for variational behaviour. Its application 
is only limited by the requirement of having functions
 strictly localised in regions of real-space 
but this is a characteristic of most modern
 real-space methods as they are designed to have a
computational cost that scales linearly with system size.
\end{abstract}


\maketitle

Electronic structure methods usually require the 
solution of  Schr\"{o}dinger's  equation 
\begin{equation}
\hat{H}\psi_0 = \varepsilon_0 \psi_0  \label{SCHR}
\end{equation}
for the ground state energy $\varepsilon_0$ and wavefunction $\psi_0$
of the Hamiltonian operator $\hat{H}$ of the system.
Exact solutions exist only in very simple cases. 
A common practical way of approximating the solution
of (\ref{SCHR}) is by expanding $\psi_0$ in terms
of a set of basis functions. It is straightforward 
to show that this \emph{basis set approach} has two 
desirable properties:
\begin{enumerate}
 \item The approximations to $\varepsilon_0$ are always
 upper bounds to the exact $\varepsilon_0$,
so any algorithm that minimises $\varepsilon_0$ with
respect to $\psi_0$ will yield the optimal solution for 
a given basis set.
 \item The leading term of the error $\delta \varepsilon_0$ 
in the energy is proportional to the square 
of the error $\delta \psi_0$ in the wavefunction and 
thus the energy displays quadratic convergence 
with increasing basis set size.
\end{enumerate}
These two properties are referred to as the \emph{variational
principle} and are characteristic of the basis set approximation.
Basis sets in common use for calculations on molecules 
and solids include Gaussian functions \cite{BOYS1,GAUSSIANS}, 
Slater functions \cite{SLATER1}, 
plane-waves \cite{IHM1979}, 
spherical Bessel functions \cite{HAYNES97}, 
pseudo-atomic orbitals \cite{SANKEY1989,SOLER2001}, 
wavelets \cite{ARIAS1999},
``blip'' cubic splines \cite{HERNANDEZ1997},
 and finite elements \cite{PASK1999}.

In recent years, the ability to perform calculations using 
only quantities that are \emph{localised in real space}, 
such as the density matrix or Wannier-type orbitals, has 
led to linear-scaling electronic structure methods that 
can deal with thousands of atoms \cite{GOEDECKER-REVIEW}. 
As a consequence, the importance of localised basis 
sets has grown because they are required for the 
efficient representation of localised functions.
Delocalised basis sets such as plane-waves
are not suitable as they make the cost of the electronic
structure calculations
 scale as the cube of the system size. 

Basis set expansion is by no means the only 
practical approximation in electronic structure theory.
A set of important alternatives are the ``real-space'' 
methods \cite{BECK-REVIEW,CHELIKOWSKY2} which 
use a real-space grid to express their solutions. 
The representation of functions directly on a real-space
grid, either regular \cite{BECK-REVIEW} or adapted to 
the positions of the atoms \cite{ACRES}, simplifies the 
application of localisation constraints 
and for this reason the importance of 
real-space methods in recent years has grown in 
parallel with that of the localised basis set methods.
Real-space methods based on a regular grid bear some 
similarity to the plane-wave pseudopotential approach as the spacing 
of the real-space grid defines  
a plane-wave kinetic energy cutoff 
and is the parameter with respect to which the 
solutions are converged.
Most real-space methods are based on the finite 
differences (FD) approach, so 
they are not variational, contrary 
to the plane-wave basis set method. 
 This is due to the fact that 
the Laplacian operator for the kinetic energy in the 
Hamiltonian is represented, or rather approximated, 
as a FD expansion 
\begin{equation}
\frac{\partial^2 \phi}{\partial x^2}(x_i, y_j, z_k)
\simeq \frac{1}{h_x^2}\sum_{n=-A/2}^{A/2}
C_n^{(A)} \phi(x_i+nh_x, y_j, z_k)  \label{FD}
\end{equation}
on every grid point $(x_i, y_j, z_k)$,  
where $h_x$ is the grid spacing in the x-direction.
Using (\ref{FD}) is equivalent to applying the exact Laplacian 
operator to an approximation of the solution 
by a polynomial of degree $A$ at each point of the 
grid \cite{ACTA_NUMERICA_94}. Therefore, the 
 Hamiltonian operator changes as the grid spacing is varied.
One consequence of this is the lack of variational
behaviour in the solution process, a well known
feature of real-space methods \cite{BECK-REVIEW}.

Maragakis \emph{et al.} \cite{MARAGAKIS2001} correctly recognised 
that the FD coefficients $C_n^{(A)}$ of equation 
(\ref{FD})  systematically underestimate 
the kinetic energy. They have  redetermined them so that they 
always slightly overestimate it, provided one assumes 
that the solution on the grid is the real-space representation of a 
plane-wave expansion. This in practice means that with their new
FD coefficients energies always converge from above as the 
grid is made finer. So at least one of the 
desirable features of the variational principle is 
restored with this amended FD approach.

A completely different approach which attempts to combine 
the merits of the plane-wave basis set method with the localisation 
properties of the real-space methods is the FFT box 
technique \cite{KINETIC_PAPER,TOTALENERGY_PAPER,NGWF_PAPER}. 
Here a plane-wave basis is assumed but the functions 
are kept localised as they are represented on a
 real-space grid from the outset. A subset of the plane-waves 
which is proportional to the number of grid points in 
the regions of the localised functions
and independent of the size of the system is used 
whenever the functions are represented in reciprocal-space.
Functions localised in real-space are smooth and delocalised 
in reciprocal-space, therefore \emph{coarse sampling} 
in reciprocal-space is a very good approximation \cite{KINETIC_PAPER}. 
There is no modification of the Hamiltonian operator
but an intermediate truncation of the 
basis set is performed, so the variational behaviour can not be 
guaranteed but can be expected as a consequence of the 
close resemblance in practice of the FFT box method to 
the full plane-wave  approach.

To assess the variational properties of the new methods,   
we will first use the exactly solvable model of the harmonic 
oscillator that Maragakis \emph{et al}. used for their tests. 
\begin{figure}

\centering

\scalebox{0.7}{\includegraphics*[0cm,1cm][20cm,28cm]{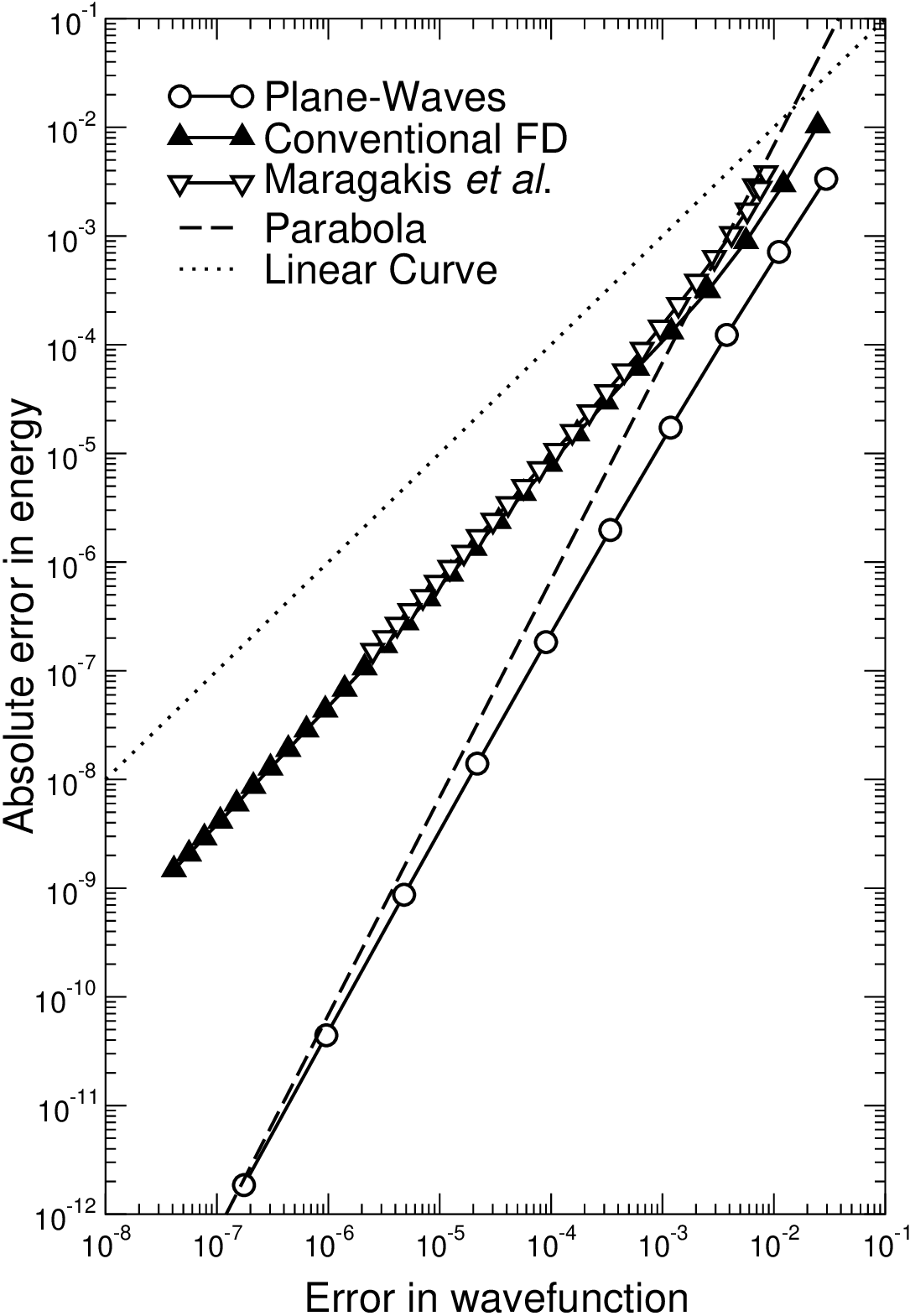} }

\caption{Error in energy as a function of the error in
the  wavefunction for the harmonic oscillator (arbitrary units).
The curves represent approximations to the ground state solution
on a real-space grid as the number of points is increased from 
16 to 18 to 20, etc. in the interval [-10, 10].  The conventional and 
Maragakis \emph{et al.}\cite{MARAGAKIS2001}
FD order $A=12$ methods were used, as well as the plane-wave
basis set expansion method (the number of 
plane-waves that can be represented on the real-space 
grid without aliasing is equal to the number of grid 
points). \label{ERRORS} }

\end{figure}
As they established in their paper that their modified 
FD method converges from above, here we examine
 if the quadratic convergence criterion 
of the variational principle is also satisfied.
We compare their method for $A$=12 with the conventional FD
 $A$=12 method and with the plane-wave expansion method which 
is variational by definition.
In Figure \ref{ERRORS} we plot 
the absolute error in the ground state 
energy as a function of the rms error in the 
ground state wavefunction. 
The behaviour of the two FD methods 
is very similar, their curves in Figure \ref{ERRORS} 
almost coincide and have a slope close to unity 
or in other words the error in the energy depends only linearly on the 
error in the wavefunction. On the other hand, the plane-wave
curve has a slope close to 2 and becomes parallel with 
the parabolic line (slope 2) as the error in the wavefunction 
decreases, so it is truly variational as expected.

\begin{figure}

\centering

\scalebox{0.7}{\includegraphics*[0cm,1cm][25cm,19cm]{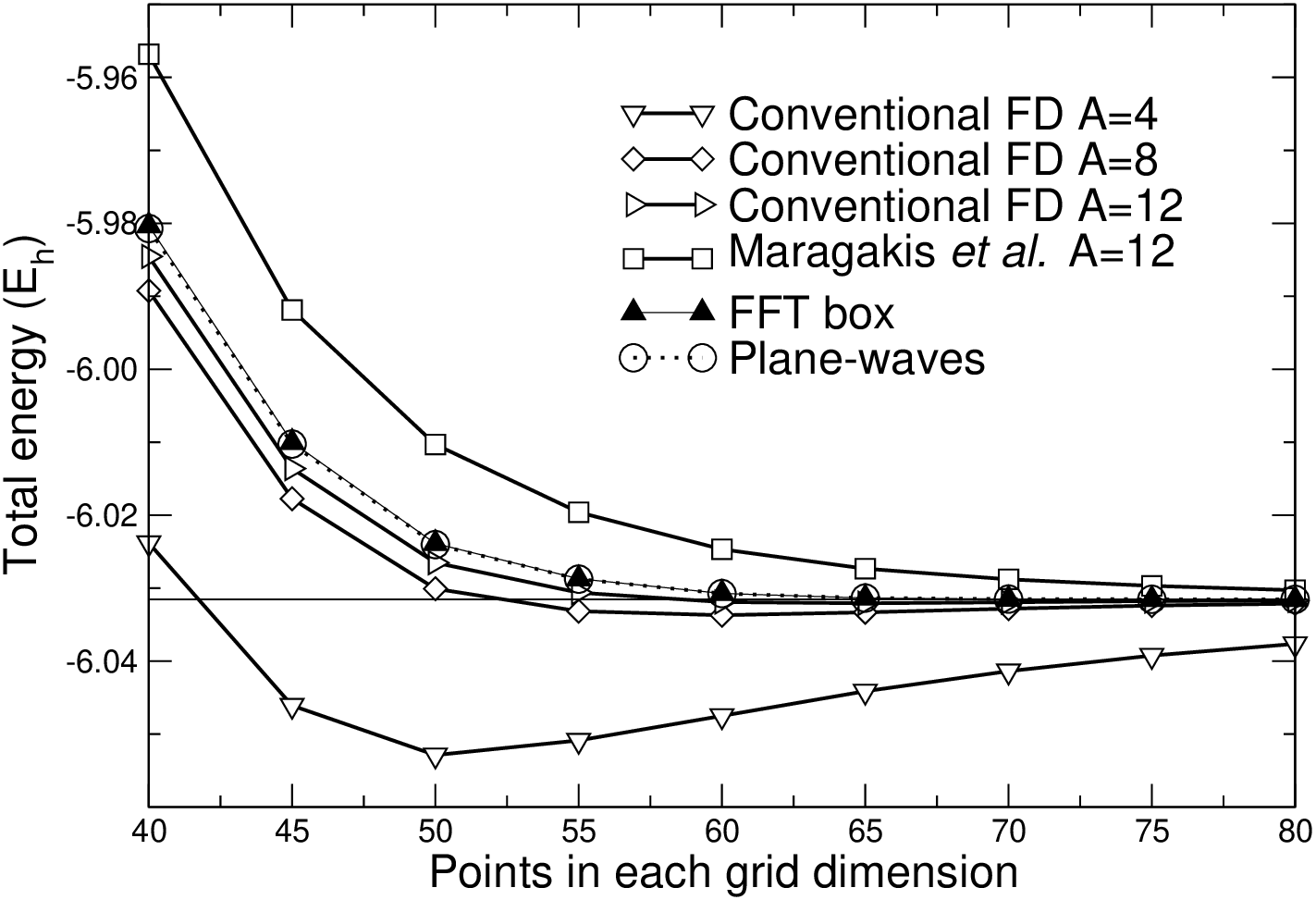} }

\caption{Convergence of the total energy of a silane molecule 
in a cubic box with respect to the number of grid points in each dimension.
The energy is calculated with conventional finite differences (FD) of 
order $A=4,$ 8 and 12, with the Maragakis \emph{et al}. \cite{MARAGAKIS2001}
variational FD of order $A=12$, with the FFT box technique \cite{KINETIC_PAPER} 
and with the 
plane-wave basis set method. The FFT box curve almost coincides
with the plane-wave curve on the scale of this graph. \label{SILANE_FDCONV} }

\end{figure}

The fact that none of the FD methods exhibit the
quadratic convergence of variational behaviour may not be a 
serious drawback for practical calculations since computational
limitations always preclude high convergence with 
respect to grid spacing. However, the speed of convergence 
is important as it will determine the speed at which
relative energy differences converge.
To examine this aspect we need to consider  an example
on a real system. For this purpose, we 
have performed LDA density-functional calculations
on a silane molecule inside a large cubic simulation cell 
of (40 \AA)$^3$. 
The atomic cores are represented by norm-conserving 
pseudopotentials \cite{TROULLIER1}
and the charge density is expressed in terms of 
pseudo-atomic orbitals \cite{SANKEY1989}
centred on the atoms and optimised for each pseudopotential
in spherical regions of 8.0 a$_0$ radius.
We compare the conventional FD methods of order $A=4, 8$ and
12 with the Maragakis \emph{et al}. variant of order $A=12$,
with the FFT box technique, and with the conventional plane-wave 
basis set approach. In all FD calculations
we solve the Poisson equation for the Hartree potential
using the conventional FD coefficients, 
as Maragakis \emph{et al}. do, so the only difference 
between the two forms of FD is in the calculation of the kinetic energy.
Figure \ref{SILANE_FDCONV} shows the convergence of the
total energy as a function 
of the number of grid points in every lattice vector 
direction.
The convergence of the energy with the conventional FD
methods is not monotonic although the effect is minimised as the 
order increases. At coarse grid spacings, FD converges 
from above since there are not enough grid points to
accurately represent the charge density near the 
ions which is the major contribution to the 
total energy. When the grid spacing becomes finer than
some threshold that depends on the order of the 
FD, the pseudopotential contribution lowers the 
energy, but to a value which is less than the 
grid-converged result beacuse of 
the systematic underestimation of the
kinetic energy.
In the case of the Maragakis \emph{et al}. FD, the convergence with 
grid spacing is monotonic and always from above, as expected 
and as we have also confirmed in tests with other systems.
In terms of the speed of convergence and deviation from the
plane-wave curve which is variational by definition, the conventional 
FDs of order $A=8$ and 12 do quite well starting with errors in the
order of m$E_h$ which decrease fast with the number of points.
The convergence of the variational FD  is not
as good on the other hand as the size of its errors is larger 
and its rate of convergence is slower.  
The FFT box technique closely follows the
plane-wave curve as it is expected to do and has errors that are 
at least an order of magnitude smaller 
than the FD $A=12$ curve throughout its range, so it apears to
coincide with the plane-wave curve on the scale of this graph.

\begin{figure}

\centering

\scalebox{0.7}{\includegraphics*[0cm,1cm][25cm,19cm]{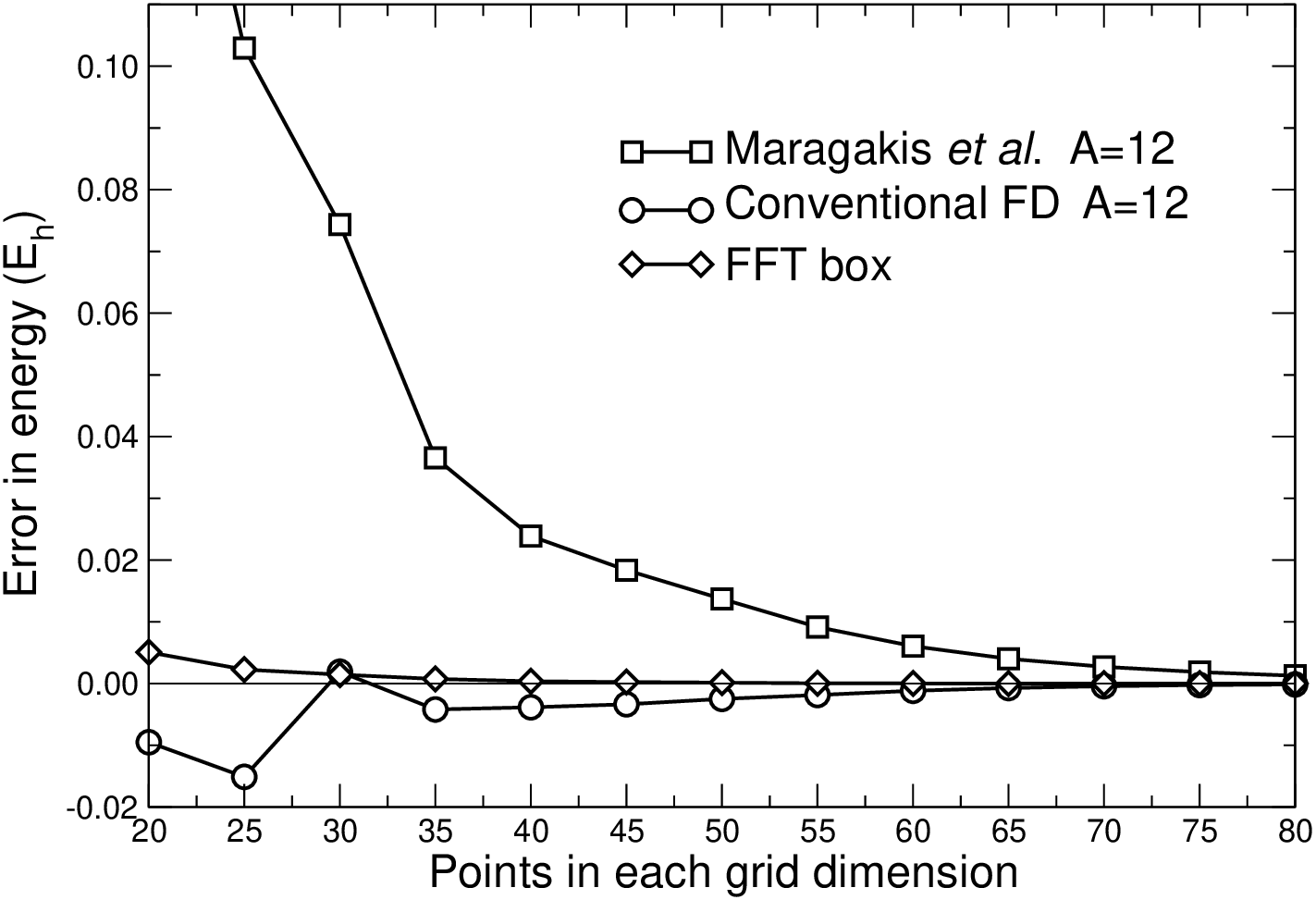} }

\caption{Convergence of the total energy of a silane molecule 
in a cubic box with respect to the number of grid points in each dimension.
The error in energy with respect to the conventional plane-wave 
calculation is plotted for the FFT box technique
and for the conventional and Maragakis et al. \cite{MARAGAKIS2001}
FDs of order $A$=12.
  \label{SILANETEST} }

\end{figure}

Some more conclusions on the behaviour of the three methods can be
derived from Figure \ref{SILANETEST}.
Here we plot the deviation of the three best cases of Figure \ref{SILANE_FDCONV}
(conventional and variational FD of $A=12$ and the FFT box 
technique) from the plane-wave curve which is variational
by definition.
The FFT box curve is closest to the basis set behaviour.
The conventional FD converges satisfactorily as long as the 
grid is dense enough but it is probably safe to say \cite{KINETIC_PAPER} 
that no matter how high an order we use, it
will always be worse than the FFT box. The FD of Maragakis 
\emph{et al}. has the desirable property of always converging 
from above but its errors are an order of 
magnitute larger than the conventional FD of the same order.

In summary, we have carried out a comparison of methods for 
electronic structure calculations directly in real-space
in order to examine if and to what extent they can 
display the variational behaviour of basis sets.
The FD representation of the Laplacian for the kinetic 
energy operator systematically underestimates the kinetic energy
and is the source of the problems as correctly 
identified by Maragakis \emph{et al}. Their proposed 
modification of FD coefficients systematically overestimates the
kinetic energy and causes the energy to converge 
from above albeit with errors up to an order of magnitude
larger than conventional FD. In practice this
means that finer grids may be necessary and
this could be a high price to pay in terms of 
computational cost. None of the FD methods display 
quadratic convergence of the energy with respect to the
wavefunction which is the second characteristic of 
the variational principle.
For cases where the functions are strictly localised 
in regions of the real-space grid, 
as with most real-space methods intended for 
linear-scaling calculations \cite{FATTEBERT2000,SOLER2001}, 
the FFT box technique could be the method of choice
 as it is designed to closely follow the behaviour 
of the fully variational plane-wave method.

C.K.S. would like to thank the EPSRC (grant number GR/M75525) for 
postdoctoral research funding. 
O.D. would like to thank the European Commission for a 
Research Training Network fellowship (grant number HPRN-CT-2000-00154).
P.D.H. would like to thank Magdalene College, Cambridge, for a 
research fellowship.

\bibliography{var_methods}

\end{document}